\documentclass[final,aps,pra,graphicx,amsmath,amssymb,
superscriptaddress]{revtex4}

\usepackage{graphicx}
\begin{document}
\title{Momentum distribution of a trapped Fermi gas with large
scattering length} 

\author{L. Viverit} 
\affiliation{Dipartimento di Fisica, Universit\`a di Trento
and BEC-INFM, I-38050 Povo, Italy}
\affiliation{Dipartimento di Fisica Universit\`a di Milano, 
via Celoria 16, I-20122 Milan, Italy}
\author{S. Giorgini}
\affiliation{Dipartimento di Fisica, Universit\`a di Trento
and BEC-INFM, I-38050 Povo, Italy}
\author{L. P. Pitaevskii}
\affiliation{Dipartimento di Fisica, Universit\`a di Trento
and BEC-INFM, I-38050 Povo, Italy}
\affiliation{Kapitza Institute for Physical Problems, 117334 Moscow, Russia}
\author{S. Stringari}
\affiliation{Dipartimento di Fisica, Universit\`a di Trento
and BEC-INFM, I-38050 Povo, Italy}

\begin{abstract}
{ Using a scattering length parametrization of the BCS-BEC
crossover as well as the local density approximation for the density profile, 
we calculate the momentum distribution of a harmonically trapped atomic Fermi
gas at zero temperature. Various interaction regimes are considered, including
the BCS phase, the unitarity limit and the molecular regime. We show that the
relevant parameter which characterizes the crossover is given by the 
dimensionless combination $N^{1/6}a/a_{ho}$, where $N$ is the number of atoms, 
$a$ is the scattering length and $a_{ho}$ is the oscillator length. 
The width of the momentum distribution is shown  to depend in a crucial way on
the value and sign of this parameter. Our predictions can be relevant
for experiments on ultracold atomic Fermi gases near a Feshbach resonance.}
\end{abstract}

\maketitle

Recent experiments on ultracold atomic Fermi gases near a 
Feshbach resonance, have 
pointed out the crucial role played by two-body interactions. 
Spectacular effects concern the hydrodynamic behaviour exhibited 
by the expansion after release of the trap \cite{thomas,regal,bourdel} 
and the evidence of molecular formation for
positive values of the scattering length \cite{jin,salomon}.
These experiments open important perspectives towards the 
realization of the superfluid phase
in Fermi gases, whose  critical temperature has been 
predicted to be significantly
enhanced by resonance effects \cite{holland,timmermans,griffin}, and
towards the study of the
BCS-BEC crossover \cite{Leggett,NS,SRE,ERS} and of many-body effects in the
presence of a Feshbach resonance \cite{combescot,bruunpeth,stoof}.

The purpose of this paper is to point out that an important 
diagnostics of the state of the
system across the resonance is the atomic momentum distribution 
whose shape, at low 
temperature, is very sensitive to the size and sign of the scattering length.
The momentum distribution can be directly measured in these 
systems either by suddenly switching off the scattering length
and imaging the expanding atomic cloud \cite{bourdel}, 
or by Bragg spectroscopy
\cite{MIT}.

In this paper we will consider a Fermi gas of two spin species. The
interspecies interaction is characterized by the s-wave scattering 
length $a$ which will be  assumed to be larger than the effective 
range $r_0$ of the interaction. Furthermore we will also assume that
the  relative momentum $\hbar k$ between the colliding atoms is smaller 
than $\bar/r_0$. If the
scattering length $a$ is large and positive a weakly bound molecule forms
in the vacuum with binding energy $E_b=-\hbar^2/ma^2$ and size 
$a$ \cite{note4}.
A peculiar feature of the resonance is that when $k|a|\gg 1$
the scattering amplitude $f$ is unitarity limited and 
$f\to i/k$, independent of the value of $a$ \cite{LanLif}.

In the many-body problem the above lengths should be compared with the
average distance between particles proportional, at $T=0$, 
to the inverse of the Fermi wavenumber 
\begin{equation}
\label{defkf}
k_F=(6\pi^2)^{1/3}n_\sigma^{1/3} \;,
\end{equation}
where the densities $n_\sigma$ of the two species will be taken equal. 
By assuming that both the average distance between
particles and the modulus of the scattering length are larger than 
the effective range $r_0$,  
the effects of interactions will not depend on the actual value of
$r_0$,  but only on the combination $k_Fa$. 

For small and negative values of $k_Fa$, a homogeneous gas at zero temperature
exhibits a BCS superfluid phase, whose momentum distribution
differs very little from the step function 
$\theta(k_F-k)$ of the non-interacting Fermi gas. 

When either the value of the scattering length or of the density
increases, correlations become more and more important. 
Eventually, in the unitarity limit 
$k_F|a| \gg 1$, the configuration of the system is expected to 
exhibit a universal behavior, independent of the  
actual value of $a$ \cite{heiselberg}.
When we move to the side of the 
Feshbach resonance where the scattering length is large and positive,
bound molecules of size $a$ would form in the vacuum.
However if $a$ is still much larger
than the interparticle spacing the actual state of the many-body system 
will be  very different. Only when $k_Fa$ becomes smaller 
than unity will these molecules not be perturbed by the medium and 
behave just as independent bosons. 
In this limit the atomic momentum distribution is expected to become 
$\sim 1/(k^2a^2+1)^2$, corresponding to 
the momentum distribution of atoms inside molecules.

In the following we shall work at $T=0$ and take into account the 
effect of the harmonic trapping. This is important 
because the actual values of the 
densities, and hence of $k_F$, realized in experiments 
depend in a crucial way on the trapping parameters. We use 
the model developed in \cite{ERS}
to investigate the BCS-BEC crossover at $T=0$, which
accounts for all the regimes introduced above 
(BCS limit, unitarity limit, BEC molecular limit) with a unique tuning 
parameter fixed, in uniform gases, 
by the combination $k_Fa$. This approach consists of a 
generalization of 
the gap equation and the number equation of the usual BCS theory 
to the whole resonance region, and corresponds to using
the saddle point approximation for the zero temperature 
partition function.

Eliminating as usual the bare interatomic potential in favor of the scattering
t-matrix in the vacuum, one obtains the result
\begin{equation}
\frac{m}{4\pi\hbar^2 a}=\frac{1}{V}\sum_{\bf k}\left(\frac{1}{2\epsilon_k}
-\frac{1}{2E_k}\right)
\label{gap}
\end{equation}
for the gap equation,where $\epsilon_k=\hbar^2k^2/2m$,
$E_k=\sqrt{(\epsilon_k-\mu)^2+\Delta^2}$, $\mu$ is the chemical potential,
$\Delta$ is the gap parameter and $V$ is the volume of the system.
The number equation, on the other hand, takes the form 
\begin{equation}
n_\sigma=\frac{1}{2V}\sum_{\bf k}\left(1-\frac{\epsilon_k-\mu}{E_k}\right) \;.
\label{number}
\end{equation}
The two equations form a closed set of equations in the unknowns 
$\Delta$ and $\mu$ and can be solved for a  given value of $k_Fa$.
The approach is reliable when 
$k_Fa\to0^\pm$ with $|a|\gg r_0$, 
and approximate in the region $k_F|a| \gtrsim 1$.

It is immediate to see that the particle distribution 
of atoms of either species is given by the expression
\begin{equation}
\label{nkgen}
n_{k}=\frac{1}{2}\left(1-\frac{\epsilon_k-\mu}{E_k}\right),
\end{equation}
and  corresponds, in the usual Bogoliubov theory of 
Fermi superfluids, to $n_k=|v_k|^2$ where $v_k$ are the 
Bogoliubov amplitudes.
The above expression for $n_k$ is only
valid for values of $k$ smaller than the inverse of the effective  range 
$r_0$. The values of $k$ can however be
much larger than $1/a$ since we assume $a\gg r_0$.

Solutions of equations (\ref{gap}) and (\ref{number}) 
are available for the whole range of values of $k_Fa$ \cite{marini,bertsch}. 
Particularly simple solutions exist in the limits
$k_Fa\rightarrow 0^\pm$.
In the BCS limit ($k_Fa\rightarrow 0^-$) the solutions are well known: 
$\mu\simeq \epsilon_F$
and $\Delta \simeq 8e^{-2}\epsilon_F \;\exp(\pi/2k_Fa)$, 
with $\epsilon_F=\hbar^2k_F^2/2m$ being the Fermi energy. 
Notice that, since the present theory 
neglects the fluctuations in the 
number-density and spin-density, the Gorkov correction to the gap $\Delta$ 
is not included \cite{gorkov,heispeth}. 
However, this has no major relevance for the calculation of the particle 
distribution which in this limit is close to the free-gas value.

In the BEC limit instead one finds
$\mu\simeq -\hbar^2/2ma^2+2\pi\hbar^2an_\sigma/m$ and 
$\Delta\simeq (16/3\pi)^{1/2}\epsilon_F/\sqrt{k_Fa}$. 
The first term in the chemical potential is
simply $E_b/2$, i.e. the molecular binding energy per atom.
By rewriting the second term as $2\pi\hbar^2a_Bn_\sigma/m_B$, 
where $a_B=2a$ and
$m_B=2m$, one can see that this accounts for
the mean field molecule-molecule interaction energy \cite{note3}.
The quantity $\Delta$ instead plays the
role of an order parameter for the molecular Bose-Einstein condensate
(see for example \cite{strinati}).

In both the BCS and BEC limits $\Delta$ is much smaller than the 
abolute value of the chemical potential,
so that the particle distribution can be expanded in terms of $\Delta/|\mu|$. 
In the BCS limit the lowest order in the expansion is simply the
free Fermi step function, and it is enough to stop at this level for the 
considerations which follow.
In the BEC limit one instead finds the result 
\begin{equation}
n_{\bf k} =  \frac{4}{3\pi}(k_Fa)^3
\frac{1}{(k^2a^2+1)^2} \;.
\label{molecule}
\end{equation}
This is proportional to the square of the Fourier transform of the 
molecular wavefunction $\psi(r)=(1/r)\;\exp(-r/a)$, indicating that in 
this limit the particle distribution arises from the motion of 
atoms inside the molecules, and the prefactor accounts for the fact that 
there are $n_\sigma$ molecules per unit volume. 

Let us now consider the most interesting  unitarity limit where 
$k_Fa \to \pm \infty$. The left hand side of Eq. (\ref{gap})
vanishes, and this reduces to an implicit 
equation for the ratio $\Delta/\mu$. A straightforward calculation 
\cite{marini,bertsch} gives $\Delta/\mu\simeq 1.16$ and from
Eq. (\ref{number}) one finally finds the results 
$\Delta\simeq 0.69\, \epsilon_F$ and
$\mu\simeq 0.59\, \epsilon_F$. 
In the unitarity limit the proportionality of the chemical potential
and of $\Delta$ with $\epsilon_F$ can be deduced from general dimensional 
arguments and follows from the fact that $a$ is no longer a relevant scale
in the problem. In this limit the chemical potential is often written in 
the form 
$\mu= (1+\beta)\, \epsilon_F$ where, in our case, $\beta\simeq -0.41$.
It is interesting to compare this prediction with the results of more
elaborated approaches.
A recent microscopic calculation based on quantum Monte-Carlo techniques
gives the value $\beta=-0.56$ \cite{carlson}, while the theory of resonance 
superfluidity based on saddle-point approximation yields the value 
$\beta\simeq - 0.35$ \cite{holland1}.
The value of $\beta$ has also been the object of recent experimental 
investigations
in trapped Fermi gases at finite temperature, through the study of the 
release energy \cite{thomas,bourdel}.

\begin{figure}[htbp]
  \begin{center}
    \scalebox{0.7}{\rotatebox{270}{
    \includegraphics[]{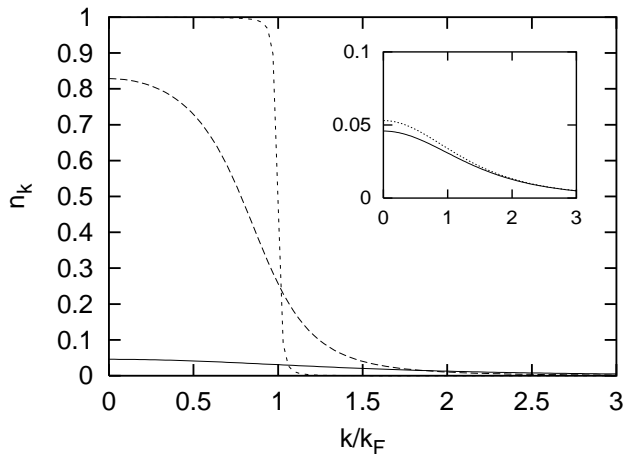}}}
    \vspace{.2cm}
    \caption{Particle distribution of a uniform gas in the
BEC (solid line), unitarity (long dashed line) and BCS regime  
(short dashed line). For the BCS regime we have chosen $k_Fa=-0.5$, and
for the BEC one $k_Fa=0.5$. In the inset the momentum distribution in the
BEC regime is compared with the molecular distribution (\ref{molecule})
(dotted line).}
    \label{nkk1}
  \end{center}
\end{figure}

The predictions of the theory for the particle distribution 
of a uniform gas are reported in Fig. \ref{nkk1}, which shows that during 
the transition
from the ideal gas (dashed line) to the molecular limit 
(full line) across the unitarity limit (long-dashed line)
the atomic momentum distribution varies in a dramatic way,
reflecting the critical role played by two-body interactions.
This behavior of the momentum distribution was first pointed out 
in \cite{NS,SRE}. More recently, it has been investigated in
\cite{griffin1} and, in the unitarity limit, by the authors of \cite{carlson}.
In particular the results of \cite{carlson} rather well agree with
the curve of Fig. \ref{nkk1} for the unitarity limit.
The width of the momentum
distribution ranges from $\,\sim k_F$ in the BCS limit to 
$\,\sim 1/a$ in the BEC regime. Notice that for the value 
$k_Fa=0.5$ reported in the figure, the momentum distribution 
calculated by solving Eqs. (\ref{gap}) and (\ref{number}), is practically 
indistinguishible from the ``molecular'' formula (\ref{molecule}). 
 
In trapped configurations 
we shall calculate the momentum distribution by introducing
the local semiclassical particle distribution
\begin{equation}
\label{locmomdist}
n_{\bf k}({\bf r})=\frac{1}{2}\left(1-\frac{\epsilon_k
-\mu({\bf r})}{E_k({\bf r})}\right),
\end{equation}
where $\Delta({\bf r})$ and $\mu({\bf r})$ are obtained by solving 
Eq. (\ref{gap}) and (\ref{number}) with the local value of the
density. The momentum distribution is then evaluated by 
integrating the particle distribution in coordinate
space:
\begin{equation}
n({\bf k}) = \int \frac{d^3 r}{(2\pi)^3}\;n_{\bf k}({\bf r}) \;.
\label{totmomdist}
\end{equation}
The momentum distribution coincides with the particle distribution
function only in a uniform system, where one has the simple relation
$n({\bf k})=V\,n_{\bf k}$. 
A fundamental ingredient for the calculation is the 
density distribution, whose shape can vary very much
depending on the regime considered and on the trapping parameters \cite{note1}. 
In the following the density
profile will be determined  using a local density 
approximation, based on the solution of the equation
\begin{equation}
\mu = \mu(n_\sigma({\bf r})) + V_{ext}({\bf r}) \;.
\end{equation}

In the deep BCS limit the particle distribution is given 
by the free distribution. For harmonic trapping one finds 
the well known results  
$n_\sigma({\bf r})=n_\sigma(0)\,[1-(r/R_{TF}^0)^2]^{3/2}$ for 
the density profile, where $R_{TF}^0=(48N_\sigma)^{1/6}a_{ho}$ is the 
Thomas-Fermi radius, and $n_\sigma(0)=(4N_\sigma/3\pi^4)^{1/2}\,a_{ho}^{-3}$
is the central density.
$N_\sigma$ is the number of fermions in each internal state,
and $a_{ho}=\sqrt{\hbar/m\omega}$ the harmonic oscillator length, with
$\omega=(\omega_x\omega_y\omega_z)^{1/3}$.
The momentum distribution takes the following form
\begin{eqnarray}
\nonumber
n({\bf k})=
 \frac{(R_{TF}^0)^3}{6\pi^2}
\left(1-\left(\frac{k}{k_F^0}\right)^2\right)^{3/2},
\label{freemomdist}
\end{eqnarray}
where the  Fermi wavenumber satisfies the relation
\begin{equation}
k_F^0a=(48)^{1/6} N_\sigma^{1/6} \frac{a}{a_{ho}} \;,
\end{equation}
and coincides with the value obtained using Eq. (\ref{defkf}) with the central
density $n_\sigma(0)$.

In the unitarity limit the density profile has the same form as 
in the ideal case, since the chemical potential in the uniform system
has the same power law dependence on the density 
($\mu \propto n^{2/3}$). The only 
difference with respect to the ideal case is an overall rescaling factor: 
$\mu = (1+\beta)\epsilon_F$, resulting in a contraction
of the Thomas-Fermi radius according to the law 
$R_{TF}^0 \to R_{TF}^0(1+\beta)^{1/4}$. 
Similarly, the central density is
rescaled according to $n(0)\to n(0)(1+\beta)^{-3/4}$ and consequently
the Fermi wavenumber, calculated in the 
center of the trap, is fixed by
\begin{equation}
k_Fa=\frac{(48)^{1/6}}{(1+\beta)^{1/4}}
N_\sigma^{1/6} \frac{a}{a_{ho}} \;.
\end{equation}
The  momentum distribution (\ref{totmomdist}) 
can then be calculated by integrating (\ref{locmomdist}),
employing the corresponding values
for $\Delta$ and $\mu$ and the rescaled density profile. 
After introducing the  dimensionless variable
$\tilde{r} = r/(1+\beta)^{1/4}R_{TF}^0$ 
one finds the result:
\begin{eqnarray}
n({\bf k})&=& \frac{(R_{TF}^0)^3
(1+\beta)^{3/4}}{4\pi^2}\int_0^1 d\tilde r \;\tilde r^2 \\
&\times&\left(1- \frac{(k/k_F^0)^2
-  (1+\beta)^{1/2}f(\tilde r)}
{\sqrt{[(k/k_F^0)^2-(1+\beta)^{1/2}f(\tilde r)]^2
+(1+\beta)^{-1}[f(\tilde r)(\Delta/\epsilon_F)]^2}}\right) \;,
\nonumber
\end{eqnarray}
where $f(\tilde r) = 1 - \tilde r^2$, $\beta=-0.41$ 
and $\Delta/\epsilon_F=0.69$.

Let us finally discuss the molecular BEC limit. 
To the extent that the condition 
$k_Fa \ll 1$ is satisfied  in the center of the trap, 
one can use the molecular distribution function
(\ref{molecule}) everywhere and the resulting momentum distribution 
will be consequently given by 
\begin{equation}
n({\bf k}) =\frac{a^3 N_\sigma}{\pi^2}\frac{1}{(k^2a^2+1)^2} \;.
\label{molectrap}
\end{equation}

Also in this case it is important to calculate the quantity $k_Fa$ 
in terms of the relevant parameters 
of the trap, and to this purpose one has to determine the 
density profile. 
The leading density dependent term $2\pi\hbar^2an_\sigma/m$
in the chemical potential arises from the  molecule-molecule mean-field
interaction. This
corresponds to the usual interaction term in the Gross-Pitaevskii theory for 
bosons interacting with scattering length $a_B=2a$.
In this limit the local density approximation yields the $T=0$ Thomas-Fermi 
profile 
\begin{equation}
\label{nmol}
n_\sigma({\bf r})=n_\sigma(0)
\left(1-\frac{r^2}{R_{TF}^2}\right),
 \end{equation}
where
\begin{equation}
n_\sigma(0)=\frac{1}{4\pi\,a\, a_{ho}^2}
\left(\frac{15 N_\sigma a}{2a_{ho}}\right)^{2/5},
\end{equation}
is the central density and $R_{TF} =a_{ho}(15N_\sigma\, a/2a_{ho})^{1/5}$
is the Thomas-Fermi radius. 
The use of the local density approximation is justified  
since the Thomas-Fermi parameter $N_\sigma a/a_{ho}$ is much 
larger than unity. 
In deriving the above expressions we have assumed that atoms and molecules 
are trapped with the same oscillator frequency $\omega$. Consequently
the boson harmonic oscillator length is given 
by $a_{ho}/\sqrt{2}$.
By evaluating the Fermi momentum at the center of the trap one finally finds
\begin{equation}
k_Fa =2.2 \left(N_\sigma^{1/6}\frac{a}{a_{ho}}\right)^{4/5}.
\end{equation}
The quantity $k_Fa$ should be sufficiently small in order to 
apply result (\ref{molectrap}) for the momentum distribution.
It is remarkable that in all of the regimes the product $k_Fa$, 
related to the gas parameter $na^3$, is fixed by the 
combination $N_\sigma^{1/6}a/a_{ho}$. This quantity 
then permits to characterize the various regimes exhibited by the 
interacting Fermi gas. The universality of this combination
is a consequence of harmonic trapping, and of the fact that
the equation of state $\mu/\epsilon_F$, evaluated in the uniform phase,  
has been assumed to be a function of  $k_Fa$. 
The above discussion also shows that the value of $k_Fa$ is rather insensitive
to the value of $N_\sigma$.

\begin{figure}[htbp]
  \begin{center}
    \scalebox{0.7}{\rotatebox{270}{
    \includegraphics[]{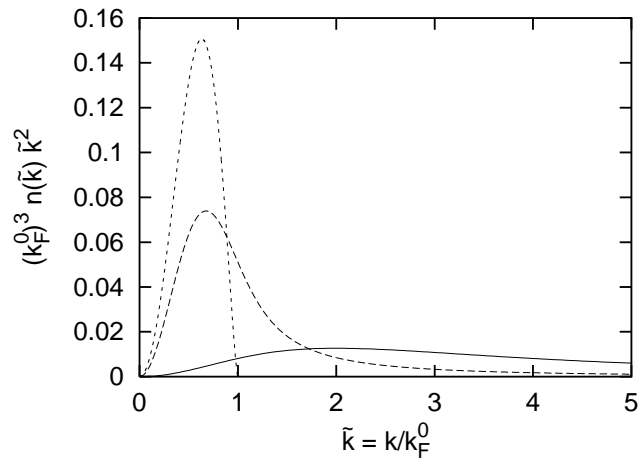}}}
    \vspace{.2cm}
    \caption{Momentum distribution of a trapped gas in the
BEC (solid line), unitarity (long dashed line) and BCS regime  
(short dashed line). In the BEC regime we have chosen
$N_\sigma^{1/6}a/a_{ho}=0.26$ corresponding to $k_Fa=0.75$ 
(and $k_F^0a=0.5$), while for the
BCS regime we have plotted the free fermion distribution. 
The momentum distributions are multiplied by
$k^2$ to emphasize the large $k$ behavior and
are normalized so that $\int\, d^3k\; n(k)=1$.}
    \label{nkk2}
  \end{center}
\end{figure}

In Fig. \ref{nkk2} we plot the momentum distribution calculated 
for three different values of $N_\sigma^{1/6}a/a_{ho}$. The 
distributions are normalized to unity and momenta are expressed in 
terms of $k/k_F^0$ where  $k_F^0$ is the Fermi wavevector of a 
non-interacting gas. Although these results are based on the approximate 
theory of \cite{Leggett,NS,SRE,ERS}, they provide a 
first useful estimate of the 
momentum distribution which might stimulate new experimental 
investigations as well as
more sophisticated and self-consistent theoretical calculations.

Let us finally discuss the experimental possibilities 
to measure the momentum distribution and to probe the behavior of the
system across the resonance.
The technique developed in \cite{salomon} to quickly turn off the scattering 
length
and to immediately release the trap is well suited for this purpose. 
If the magnetic field is
switched off in times shorter than the inverse of the binding energy
$\hbar^2/ma^2$, the molecules are suddenly dissociated and the atoms will
expand ballistically. 
If instead the scattering length is switched off adiabatically
the system will form deeper molecular states characterized by
microscopic sizes.
The imaging of the expanding atomic cloud provides a direct measurement of the
momentum distribution of the initial correlated configuration.
An important feature emerging from our results 
(see Figs. \ref{nkk1} and \ref{nkk2}) is the presence of
large $k$ components in the momentum distribution, with the consequent
suppression of $n(k)$ at small $k$. 
These large $k$ components, whose presence
is more and more pronounced
as one leaves the ideal Fermi gas condition $k_Fa \to 0^-$, affect sizably
the release energy
of the  atomic cloud  fixed by the kinetic energy \cite{note2}.

If one instead releases the trap by keeping the scattering length on, the
scenario of the expansion will
be completely different. In this case one expects that the expansion will
be governed by the laws of hydrodynamics. This will be the case both
at zero temperature, where the system is superfluid, and above $T_c$ if
$k_F|a|\gg 1$ due to the resonant effect in the collisional cross section.
Under these conditions 
the expansion will be anisotropic \cite{menotti} if the confining trap is
not symmetric and the release energy, as well as the shape of the
expanding cloud, will be determined by the
equation of state of the gas. In the unitarity limit the release energy is
given by $E_R/N_\sigma= (3/8)(1+\beta)^{1/2}(6N_\sigma)^{1/3}\hbar\omega$,
while in the molecular regime
the expansion will be similar to the one of a dilute Bose gas and one
predicts $E_R/N_\sigma=(1/7)(15N_\sigma a/a_{ho})^{2/5}\hbar \omega$. 

So far the discussion has been restricted to zero temperature. An 
interesting question concerns the behavior of the momentum distribution
at finite temperatures. For temperatures just
above $T_c$, the momentum distribution can be determined through the 
calculation of the single particle Green's function by summing the 
particle-particle ladder diagrams along the lines of \cite{NS}. 
In particular,
for negative scattering length and $k_F|a|\ll 1$, 
where $k_BT_c\ll \epsilon_F$,  one finds that the 
momentum distribution of the free Fermi gas is slightly broadened due to 
thermal
effects. In the opposite molecular
regime ($a>0$ and $k_Fa\ll 1$) one recovers the momentum distribution  
(\ref{molecule}) of atoms inside molecules, provided the binding energy of the 
molecule is much larger than the temperature $\hbar^2/ma^2 \gg k_BT$.
In fact at the critical temperature for the Bose-Einstein 
condensation of molecules
one finds $k_BT/|E_B|=0.22\,(k_Fa)^2$, and the thermal dissociation of
molecules is negligible if $k_Fa$ is sufficiently small. 

In conclusion, in this paper we have reported a first calculation of the 
momentum distribution 
of a harmonically trapped two-component Fermi gas as a function of the 
scattering length in the BCS-BEC crossover. We expect that the large 
deviations from the ideal Fermi gas distribution can be addressed 
experimentally in systems close to a Feshbach resonance.

We thank A. Perali for useful comments. 

\begin{thebibliography}{20}
\bibitem{note4} Recent experiments on $^{85}$Rb and $^{40}$K show that 
the binding energy of the molecular configurations is only semi-quantitatively
provided by the law $-\hbar^2/ma^2$, unless one works close to the
resonance, and hence that a more complete description
of the two-body problem is in general required.
\bibitem{note2} The value of the kinetic energy of the 
atomic cloud cannot be  calculated within the present
approach since $n(k)$  decreases like $1/k^4$ for large $k$ and consequently
the integral $\int d^3k\,k^2n(k)$ diverges. This unphysical divergence can
be understood recalling that Eq. (\ref{nkgen}) is only correct for 
$k\ll 1/r_0$.
\end{thebibliography}
\end{document}